\documentclass{article}

\usepackage{microtype}
\usepackage{graphicx}
\usepackage{array,caption}
\usepackage{subfigure}
\usepackage{booktabs} 
\usepackage{amsmath,mathtools}
\usepackage{enumitem}
\usepackage[dvipsnames]{xcolor}
\usepackage{todonotes}
\usepackage{tikz}
\usetikzlibrary{positioning, shapes.geometric, shapes.misc, calc}

\usepackage[hyphens]{url}

\usepackage{hyperref}
\usepackage[capitalize]{cleveref}

\usepackage[accepted]{icml2021}

\expandafter\def\csname cvxname0\endcsname{Normal}
\expandafter\def\csname cvxname1\endcsname{Classic}
\expandafter\def\csname cvxname2\endcsname{Indet}
\expandafter\def\csname cvxname3\endcsname{Other}
\newcommand{\CVX}[1]{\textsc{\csname cvxname#1\endcsname}}
\newcommand{\Sugg}{\textsc{Sugg}}
\newcommand{\given}{\,|\,}

\begin{document}

\twocolumn[
\icmltitle{Hierarchical Analysis of Visual COVID-19 Features from Chest Radiographs}




\begin{icmlauthorlist}
\icmlauthor{Shruthi Bannur}{msrc}
\icmlauthor{Ozan Oktay}{msrc}
\icmlauthor{Melanie Bernhardt}{msrc}
\icmlauthor{Anton Schwaighofer}{msrc}
\icmlauthor{Rajesh Jena}{msrc}
\icmlauthor{Besmira Nushi}{msrr}
\icmlauthor{Sharan Wadhwani}{uhb}
\icmlauthor{Aditya Nori}{msrc}
\icmlauthor{Kal Natarajan}{uhb}
\icmlauthor{Shazad Ashraf}{uhb}
\icmlauthor{Javier Alvarez-Valle}{msrc}
\icmlauthor{Daniel C.\ Castro}{msrc}
\end{icmlauthorlist}

\icmlaffiliation{msrc}{Microsoft Research, Cambridge, UK}
\icmlaffiliation{msrr}{Microsoft Research, Redmond, USA}
\icmlaffiliation{uhb}{University Hospitals Birmingham, UK}

\icmlcorrespondingauthor{Daniel C.\ Castro}{dacoelh@microsoft.com}


\vskip 0.3in
]



\printAffiliationsAndNotice{}  

\begin{abstract}
Chest radiography has been a recommended procedure for patient triaging and resource management in intensive care units (ICUs) throughout the COVID-19 pandemic.
The machine learning efforts to augment this workflow have been long challenged due to deficiencies in reporting, model evaluation, and failure mode analysis.
To address some of those shortcomings, we model radiological features with a human-interpretable class hierarchy that aligns with the radiological decision process.
Also, we propose the use of a data-driven error analysis methodology to uncover the blind spots of our model, providing further transparency on its clinical utility.
For example, our experiments show that model failures highly correlate with ICU imaging conditions and with the inherent difficulty in distinguishing certain types of radiological features.
Also, our hierarchical interpretation and analysis facilitates the comparison with respect to radiologists' findings and inter-variability, which in return helps us to better assess the clinical applicability of models.
\end{abstract}

\section{Introduction}\label{sec:intro}
Patients affected with COVID-19 frequently experience an upper respiratory tract infection or pneumonia that can rapidly progress to acute respiratory failure, multiple organ failure and death \citep{zhou2020wuhan}. Chest radiography (chest X-ray; CXR) is a front-line tool that is used in screening and triaging varieties of pneumonia due to the diagnostic role of imaging features \citep{toussie2020features} and its quick turnaround time \citep{wong2020findings}, which makes it convenient for patient management in intensive care units. However, in public healthcare systems (e.g.\ UK NHS) it can take several hours from CXR acquisition until a reporting radiologist is available to provide such a reading or RT-PCR test results become available. Thus, in this clinical context, it is especially relevant to build automated CXR image analysis systems that can benefit front-line patient management processes to decide on the clinical pathway for each patient by providing radiological feedback at point of care. This way, hospitals can better allocate resources and reduce COVID-19 contamination risks before other clinical data (e.g.\ lab tests) is made available.

\begin{figure}[t]
    \centering
    \subfigure[`\CVX0']{
        \includegraphics[width=.47\linewidth]{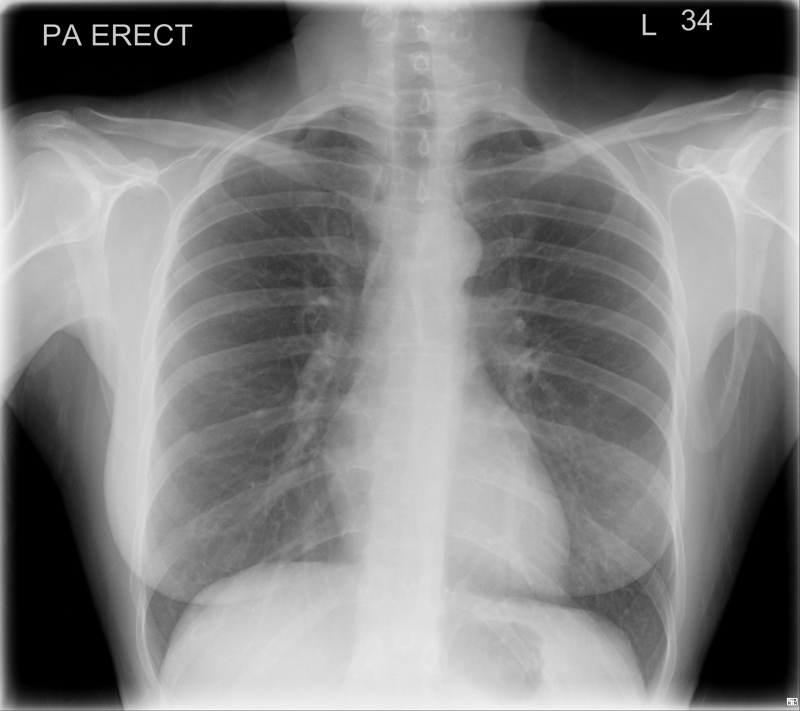}}
    \subfigure[`\CVX1' COVID-19]{
        \includegraphics[width=.47\linewidth,trim={0 8.7cm 0 0},clip]{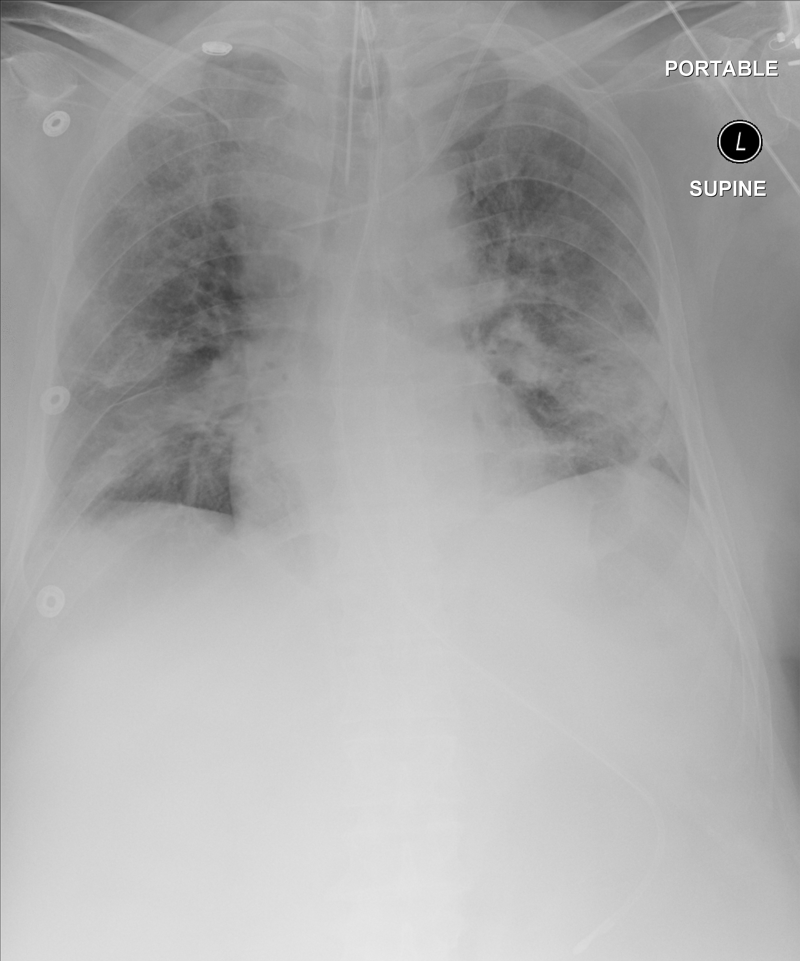}} \\
    \subfigure[`\CVX2.' for COVID-19]{
        \includegraphics[width=.47\linewidth]{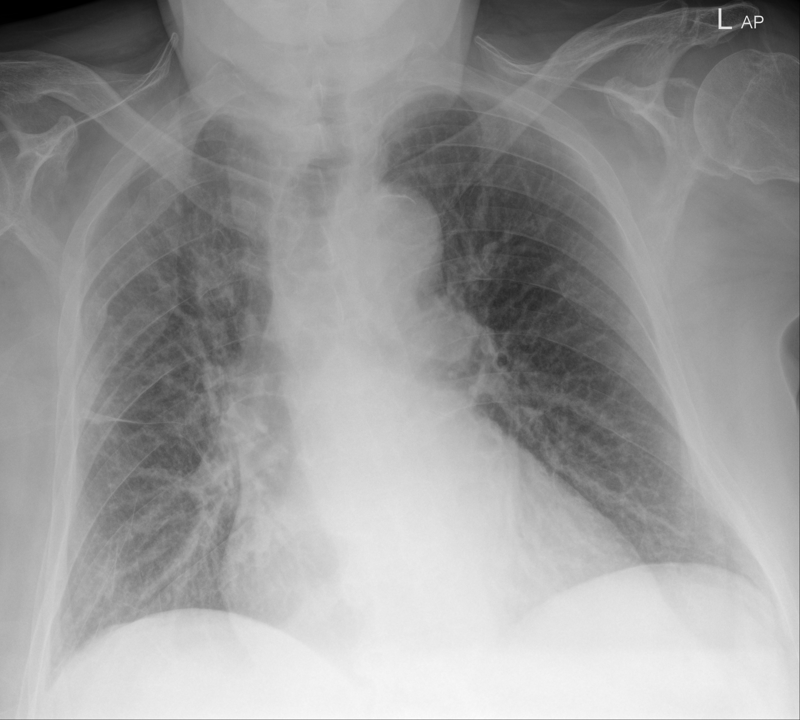}}
    \subfigure[`\CVX3' abnormality]{
        \includegraphics[width=.47\linewidth, trim={2.73cm 0 2.73cm 0},clip]{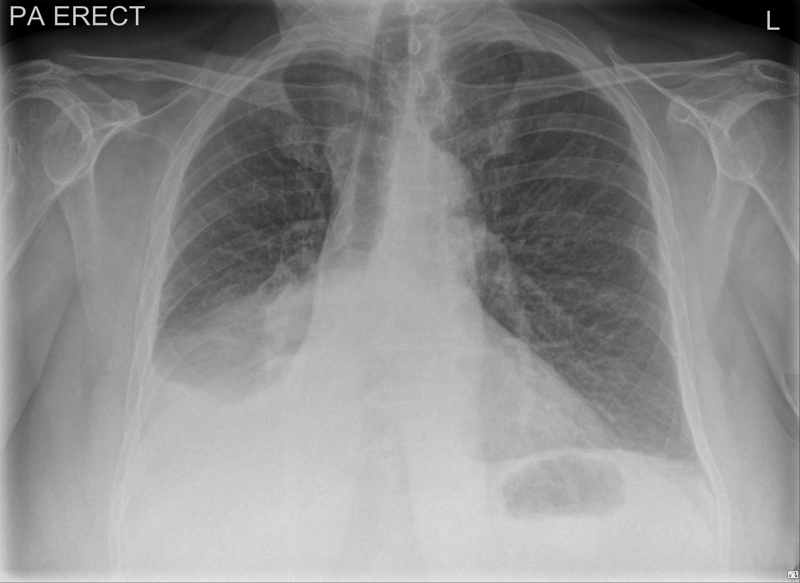}} \\
    \vskip -0.1in
    \caption{Example chest X-rays from each category. 
    (a)~Normal lungs appear mostly black.
    (b)~Bilateral, peripheral opacities (white `clouds') in lower lung lobe.
    (c)~Non-typical COVID features, with peripheral coarse white lines.
    (d)~Here, a right-sided basal pleural effusion (fluid accumulation in the left of the image).
    Images reproduced with permission.}
    \label{fig:cxr_examples}
\end{figure}

There have been several efforts to leverage machine learning models to automate this CXR reading process, reviewed in \citet{wynants2020review}. For example, \citet{wehbe2020deepcovidxr} benchmarked deep networks on a multi-site dataset comprised of thousands of CXR scans, where the models were trained to predict RT-PCR test results. Instead, in this study, we aim to predict a radiologist's impression of a chest image by identifying radiological features associated with COVID-19 pneumonia, rather than lab results or patient outcomes. CXR alone is known not to have enough diagnostic power for predicting RT-PCR results \citep{cleverley2020role}, hence the proposed tool is \emph{not} intended for diagnosis or prognosis on its own. Rather, it provides supporting evidence in addition to other readings and test results.

The structure of model outputs is directly informed by radiologists' decision-making process and was developed in close collaboration with a hospital trust, in contrast to past COVID-19 ML studies performed without clinical involvement \citep{tizhoosh2021covid}.
In detail, we propose a post-hoc interpretation of model outputs representing the clinically meaningful hierarchical relationships between target classes, as illustrated in \cref{fig:hierarchical_model} \citep[cf.][]{chen2019hierarchical}.

Our study also focuses on the potential issues and varying practices around model evaluation and biases across different hospital settings \citep{roberts2021pitfalls}. Such pitfalls can cast a doubt on the clinical applicability of the models evaluated in previous studies. As a step in this direction, we propose the use of a dedicated error analysis methodology \cite{nushi2018accountable} to understand and communicate the dependency between model's failure modes and sample attributes, such as sample difficulty and image acquisition settings. In this way, users can be made aware of reliable operating regimes of such models and ensure clinical safety whilst offering actionable solutions to address such issues.

\section{Methodology}

\subsection{Label definitions}
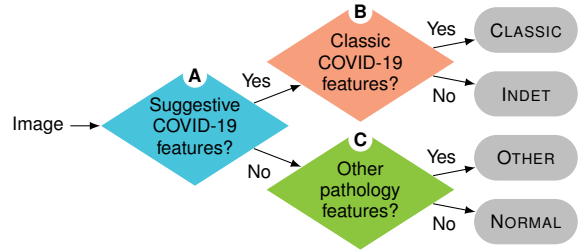
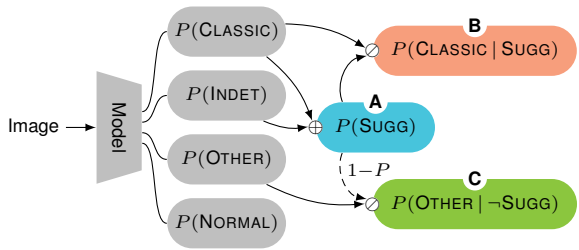
\begin{figure}
\centering%
\colorlet{color1}{SkyBlue}%
\colorlet{color2}{Melon}%
\colorlet{color3}{LimeGreen}%
\colorlet{color0}{lightgray}%
\tikzset{x=22mm, y=8.5mm, on grid}%
\tikzstyle{decision} = [diamond, draw=none, fill=color0, align=center, minimum height=16mm, minimum width=25mm, inner sep=-10pt]%
\tikzstyle{terminal} = [rounded rectangle, draw=none, fill=color0, align=center, minimum height=6mm, minimum width=18mm, inner sep=6pt]%
\tikzstyle{tag} = [circle, draw=none, fill=white, inner sep=0pt, minimum width=3.2mm, font=\bfseries]%
\subfigure[Radiologists' decision process]{
    \label{fig:decision_tree}
    \centering%
    \scriptsize%
    \sffamily%
    \begin{tikzpicture}
        \node[decision, fill=color1] (cvx12vs03) {Suggestive\\COVID-19\\features?};
        \node[decision, above right=1 and 1 of cvx12vs03, fill=color2] (cvx1vs2) {Classic\\COVID-19\\features?};
        \node[decision, below right=1 and 1 of cvx12vs03, fill=color3] (cvx0vs3) {Other\\pathology\\features?};
        \node[terminal, above right=.5 and 1 of cvx1vs2] (cvx1) {\CVX1};
        \node[terminal, below right=.5 and 1 of cvx1vs2] (cvx2) {\CVX2};
        \node[terminal, above right=.5 and 1 of cvx0vs3] (cvx3) {\CVX3};
        \node[terminal, below right=.5 and 1 of cvx0vs3] (cvx0) {\CVX0};
        \node[left=4mm of cvx12vs03.west] (image) {Image};
        
        \draw (cvx12vs03.north) node[tag, yshift=-4pt] {A}
            (cvx1vs2.north) node[tag, yshift=-4pt] {B}
            (cvx0vs3.north) node[tag, yshift=-4pt] {C};
        
        \draw[-latex] (image) edge (cvx12vs03)
            (cvx12vs03) edge node[midway, above left] {Yes} (cvx1vs2) edge node[midway, below left] {No} (cvx0vs3)
            (cvx1vs2) edge node[pos=.7, above left] {Yes} (cvx1) edge node[pos=.7, below left] {No} (cvx2)
            (cvx0vs3) edge node[pos=.7, above left] {Yes} (cvx3) edge node[pos=.7, below left] {No} (cvx0);
    \end{tikzpicture}} \\
\subfigure[Hierarchical interpretation of multi-class predictions]{
    \label{fig:model_arch}
    \centering%
    \scriptsize%
    \sffamily%
    \begin{tikzpicture}
        \tikzstyle{net} = [trapezium, trapezium angle=60, trapezium stretches=true,
                           rotate=270, draw=none, fill=color0,
                           minimum height=6mm, minimum width=15mm, inner sep=0pt]
        \tikzstyle{terminal} += [minimum width=20mm]

        \node[net] (model) {Model};
        \draw (.65,1.5) node[terminal] (cvx1) {$P(\CVX1)$}
            ++(0,-1) node[terminal] (cvx2) {$P(\CVX2)$}
            ++(0,-1) node[terminal] (cvx3) {$P(\CVX3)$}
            ++(0,-1) node[terminal] (cvx0) {$P(\CVX0)$};
        \draw (1.55,0) node[terminal, fill=color1] (cvx12vs03) {$P(\Sugg)$}
            +(.6, 1.2) node[terminal, fill=color2] (cvx1vs2) {$P(\CVX1 \given \Sugg)$}
            +(.6,-1.2) node[terminal, fill=color3] (cvx0vs3) {$P(\CVX3 \given \neg\Sugg)$};
        \node[left=4mm of model.south] (image) {Image};
        
        \draw (cvx12vs03.north) node[tag] {A}
            (cvx1vs2.north) node[tag] {B}
            (cvx0vs3.north) node[tag] {C};
        
        \path[-latex] (image) edge (model);
        \path[out=0, in=180, -]
            ($(model.north)+(0, 6pt)$) edge[looseness=.5] (cvx1)
            ($(model.north)+(0, 2pt)$) edge (cvx2)
            ($(model.north)+(0,-2pt)$) edge (cvx3)
            ($(model.north)+(0,-6pt)$) edge[looseness=.5] (cvx0);
        
        \tikzstyle{op}=[circle, fill=white, inner sep=-1pt]
        \draw (cvx12vs03.west) node[op] (m1) {$\oplus$}
            (cvx1vs2.west) node[op] (m2) {$\oslash$}
            (cvx0vs3.west) node[op] (m3) {$\oslash$};
        
        \path[-latex] (cvx1) edge[bend left=15] (m1)
            (cvx2) edge[bend right=15] (m1)
            (cvx1) edge[bend left=15] (m2)
            (cvx3) edge[bend right=15] (m3)
            (cvx12vs03) edge[bend left=50] (m2)
            (cvx12vs03) edge[densely dashed, bend right=50] node[right, pos=.3] {\scriptsize$1{-}P$} (m3);
    \end{tikzpicture}
}
\vskip -0.1in
\caption{Clinically informed model interpretation. The model's outputs are mapped to the radiologists' decision branches (A/B/C); branch probabilities are computed according to \cref{eq:aggregation}.}
\label{fig:hierarchical_model}
\end{figure}

The prediction targets were adapted from the British Society for Thoracic Imaging (BSTI)'s reporting guidelines for COVID-19 findings from chest radiographs \citep{bsti2020guidelines}:
\begin{itemize}[nosep, parsep=3pt]
    \item \CVX0: No radiological features of COVID-19 nor of other abnormalities.
    \item \CVX1: Stereotypical presentation, e.g.\ lower lobe, peripheral predominant, multifocal, bilateral opacities.
    \item \CVX2\textsc{(erminate)}: Findings that are compatible with COVID-19 presentation but nonspecific.
    \item \CVX3: Abnormalities that are not suggestive of COVID-19, e.g.\ pneumothorax, lobar pneumonia, pleural effusion, pulmonary oedema, etc.
\end{itemize}
Examples are shown in \cref{fig:cxr_examples}.
The labelling process of our partner radiologists is visualised in \cref{fig:decision_tree} and can be summarised as follows: if there are any visual features suggestive of COVID-19, the decision is narrowed to \CVX1 vs \CVX2---even if the image presents other findings like heart failure.
In other words, an image is labelled \CVX3 only if it shows no signs of COVID-19. Note that these categories refer exclusively to the presented radiological features, and labels are assigned with no access to other relevant information such as the patient record or past scans.

\subsection{Hierarchical analysis of COVID-19 features}
\def\PA{P(\Sugg)}
\def\PB{P(\CVX1\given\Sugg)}
\def\PC{P(\CVX3\given\neg\Sugg)}
To enable analysing model outputs at clinically meaningful levels of abstraction, we propose to hierarchically aggregate class probabilities reflecting the radiologists' decision process (\cref{fig:hierarchical_model}).
Unlike the related work of \citet{chen2019hierarchical}, who modelled a taxonomy of (non-COVID) thoracic abnormalities with a hierarchical model architecture, we focus on post-hoc interpretation of a vanilla multi-class predictor.


Specifically, the model is a conventional CNN classifier trained with cross-entropy loss (details in \cref{sec:implementation_details}).
We then use the four-class model outputs to compute normalised binary probabilities for each of the decision branches:
\begin{equation}\label{eq:aggregation}
    \footnotesize
    \mathclap{\begin{aligned}
    \PA &= P(\CVX1) + P(\CVX2) \,, \\
    \PB &= P(\CVX1) / \PA \,, \\
    \PC &= P(\CVX3) / (1-\PA) \,, \\
    \end{aligned}\hphantom{X}}
\end{equation}
where `\Sugg' refers to `suggestion of COVID-19' (i.e.\ `\CVX1 or \CVX2').



\subsection{Self-supervised pre-training}
The scarcity of large, annotated datasets in chest radiology has been one of the major drivers for leveraging self-supervised model pre-training \citep{sriram2021covid}. As self-supervision does not require image labels, it significantly reduces the expert annotation burden and enables local clinical sites to build on pre-trained models using much smaller private datasets.
Further, because the external datasets' labels that may not align with the target task, self-supervised learning partly mitigates widespread issues in COVID-19 studies around inappropriate ground-truth labels and misuse of external datasets \citep{roberts2021pitfalls}.

In our study, we employ the BYOL algorithm \citep{grill2020byol} and experiment with two large chest X-ray datasets for model pre-training: NIH-CXR \citep{wang2017nih}, with 112,120 frontal-view CXR scans, and CheXpert \citep{irvin2019chexpert}, with 224,316.
Both datasets were acquired from large cohorts of subjects diagnosed with chest pathologies including consolidation, pneumonia, lung nodules, etc.
Even though these external datasets contain only pre--COVID-19 scans, self-supervision enables the model to learn generic characteristics of lung lobes, opacities, and nodules, which are useful in quantifying COVID-19--related features.



\section{Datasets}

\textbf{COVID-19 CXR dataset:} 
The labelled images used for supervised fine-tuning come from a retrospective dataset comprising de-identified chest radiographs, collected in the UK across multiple sites of the University Hospitals Birmingham NHS Foundation Trust during the first COVID-19 wave (1st March to 7th June 2020). The study participants were consecutive patients who had CXR taken for suspected COVID-19 infection, presented in the emergency department, acute medical unit, or inpatient unit.
The initial dataset with 6125 images was curated by excluding duplicates and poor-quality scans, resulting in 4,940 usable images (\CVX0:~1154, \CVX1:~1778, \CVX2:~1093, \CVX3:~915).
Of the 3639 unique subjects in the curated dataset, 515 (14.2\%) were aged below 40, 2931 (80.5\%) were in the 40--89 age group, and the remaining 193 (5.3\%) were over 90; 1622 (44.6\%) were female patients.

Class labels were extracted from reports by consultant radiologists or specialist radiographers, then blindly reviewed by a radiologist based on the images alone to mitigate biases due to availability of clinical side-information. Cases for which the assigned label disagreed with the original reports were further reviewed by a clinician to reach a consensus.
Although not used as a prediction endpoint, RT-PCR test status at imaging time was also recorded for evaluation.

\textbf{Multi-label test dataset:}\label{sec:multi_label_dataset} A subset of this data was held out from training to evaluate the model's predictive performance against a diverse panel of front-line radiology reporters, in an inter-observer variability (IOV) study. It contains 400 images acquired in April 2020 (approx.\ 100 consecutive patients from each of the four categories), with age and gender distributions similar to the training set.
This test dataset was labelled into the four categories separately by three annotators with varying levels of experience in chest radiology: a consultant (\textit{Ann.~1}) and a trainee chest radiologists (\textit{Ann.~2}), and a non-specialist clinician (\textit{Ann.~3}). They had access only to these 400 images, without any other clinical context.

\section{Results}
A DenseNet-121 backbone \citep{huang2017densenet} was first pre-trained with self-supervision on NIH-CXR (cf.\ comparison with CheXpert in \cref{app:additional_results}), then fine-tuned on the private COVID-19 training dataset with 5-fold cross-validation. It was ensured that all images from the same patient were either in the training or in the validation set (see \cref{sec:implementation_details}).
The results discussed below are based on an ensemble of the 5 trained models for the best hyperparameter configuration.
Our open-source implementation is available at \url{https://aka.ms/innereyeoss}.

\subsection{Classification performance}
\begin{table}[t]
    \centering
    \vskip -0.1in
    \caption{Test accuracies of the model and clinicians ($N=400$; \CVX0: 100, \CVX1: 101, \CVX2: 98, \CVX3: 101; $\kappa$:~Fleiss' kappa statistic, indicating inter-annotator agreement)}
    \label{tab:iov_results}
    \vskip 0.1in
    \setlength{\tabcolsep}{3pt}
    \footnotesize
    \begin{tabular}{@{}l>{\bf}cccc>{(}c<{)}@{}}
        \toprule
        Classification task & \multicolumn{1}{c}{Model} & Ann.~1& Ann.~2& Ann.~3& $\kappa$ \\
        \midrule
        \textsc{Sugg.\ COVID}&.800  & .775  & .730  & .700  & .491 \\
        \CVX1 vs \CVX2      & .724  & .608  & .482  & .588  & .245 \\
        \CVX0 vs \CVX3      & .791  & .731  & .706  & .721  & .488 \\
        \midrule
        Multi-class         & .588  & .563  & .463  & .488  & .408 \\
        \bottomrule
    \end{tabular}

\vskip\floatsep
    \includegraphics[width=\linewidth]{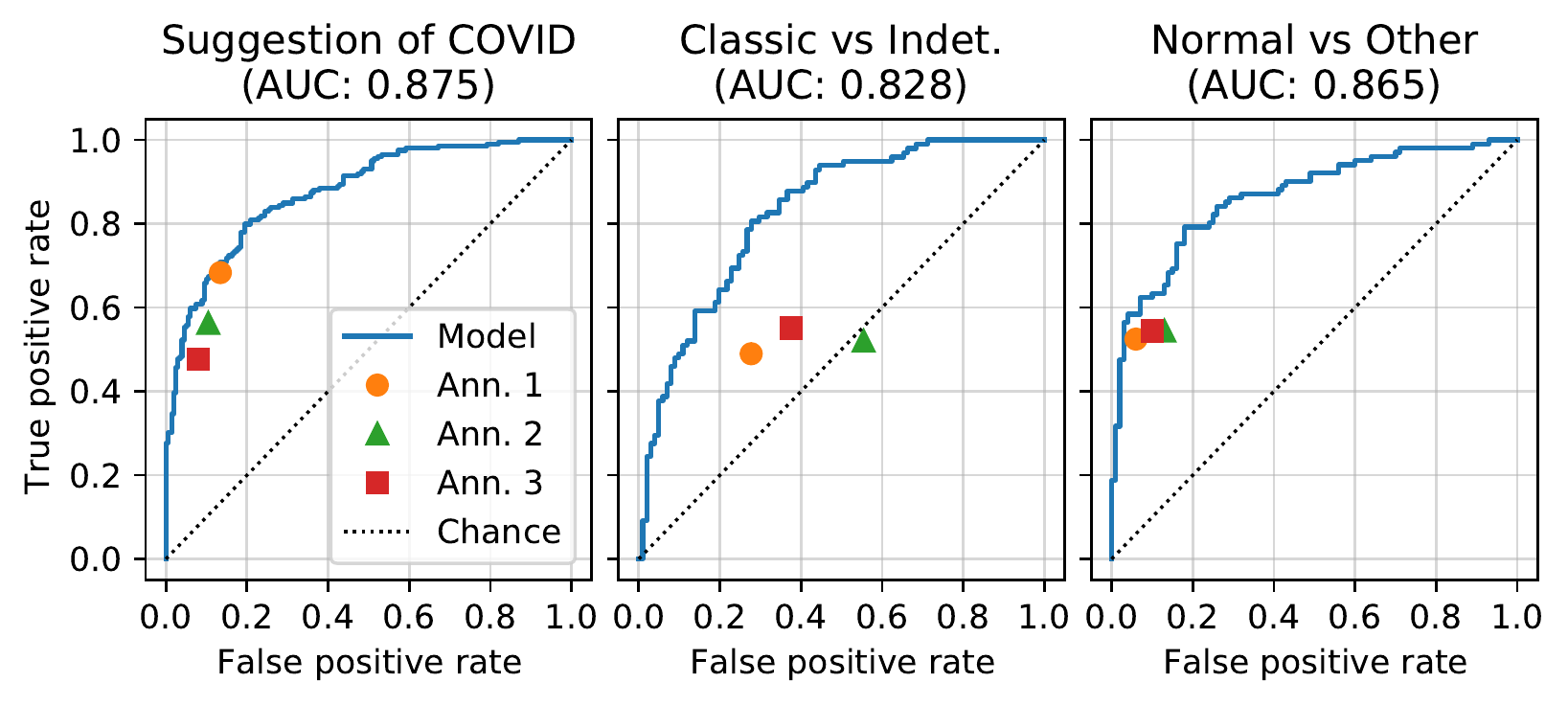}%
    \captionof{figure}{ROC curves for hierarchically aggregated model predictions, compared to clinicians' performance}
    \label{fig:roc_curves}
    \vskip -0.1in
\end{table}

\Cref{tab:iov_results} presents the test accuracies for multi-class and for each binary classification branch illustrated in \cref{fig:hierarchical_model} (A, B, and C). The corresponding ROC analysis of these binary sub-tasks is shown in \cref{fig:roc_curves}.
Reported results for `\CVX1 vs \CVX2' and `\CVX0 vs \CVX3' include only images with the relevant labels.

We see that the model outperforms the clinicians across all hierarchical and multi-class tasks, with respect to the reference labels defined as in the training dataset.
Our model achieves lower scores on `\CVX1 vs \CVX2', seemingly the hardest sub-task of the three, and the clinicians perform closer to chance level.
However, this is an inherently ambiguous problem---discussions with the annotators (including the original labellers) revealed they tended to use different thresholds for distinguishing these two classes, which is also reflected in the lowest $\kappa$ value (\cref{tab:iov_results}).

\subsection{Model failure analysis}

To better understand the model's failure patterns, we employed a semi-automatic error analysis tool \citep{nushi2018accountable} that trains a decision tree to identify partitions of data on which the model underperforms, according to attributes that are most predictive of mistakes.
Such a principled analysis of prediction errors can significantly benefit predictive models in healthcare settings by identifying potential biases and opportunities for model improvement.

The attributes considered in our analysis (see \cref{fig:error_analysis}) included RT-PCR test status for SARS-CoV-2 (positive, negative, or unknown) and X-ray acquisition direction (posteroanterior or anteroposterior view; PA/AP).
For this analysis, we focused on the top-level binary classification task (`\textsc{Sugg.\ COVID}'), which is the most relevant for patient management in a hospital. The analysis identified that the acquisition direction had the strongest association with model errors; in particular, AP images showed a higher error rate than PA.
This is consistent with the clinical context, as PA imaging is used as standard-of-care with higher diagnostic quality, whereas AP is reserved for cases when the patient is too ill to stand upright---correlating strongly with being in ICU and presenting much higher variations in image appearance and layout of the patient anatomy. In addition, AP scans with negative or unknown RT-PCR status display elevated error rates, conceivably due to the higher prevalence of \CVX2 and \CVX3 in this group.

\begin{figure}
    \centering
    \newcommand{\formatnode}[3]{#1\,/\,#2 \textit{\color{gray}(#3\%)}}
    \tikzstyle{treenode}=[rectangle, fill=black!10]
    \tikzstyle{level 1}=[level distance=25mm, sibling distance=18mm]
    \tikzstyle{level 2}=[level distance=30mm, sibling distance=6mm]
    \tikzstyle{el}=[fill=white, fill opacity=.7, text opacity=1, inner sep=2pt]
    \tikzstyle{highlight}=[fill=red!20]
    \begin{tikzpicture}[align=center, grow=right, sloped]
        \scriptsize
        \sffamily
        \node[treenode] {\formatnode{80}{400}{20.0}}
            child { node[treenode, highlight] {\formatnode{62}{278}{22.3}}
                    child { node[treenode, highlight] {\formatnode{19}{75}{25.3}}
                            edge from parent node[el, below] {PCR--} }
                    child { node[treenode, highlight] {\formatnode{14}{48}{29.2}}
                            edge from parent node[el] {PCR?} }
                    child { node[treenode] {\formatnode{29}{155}{18.7}}
                            edge from parent node[el, above] {PCR+} }
                    edge from parent node[below, pos=.4] {AP view} }
            child { node[treenode] {\formatnode{18}{122}{14.8}}
                    child { node[treenode] {\formatnode{4}{22}{18.2}}
                            edge from parent node[el, below] {PCR--} }
                    child { node[treenode] {\formatnode{8}{70}{11.4}}
                            edge from parent node[el] {PCR?} }
                    child { node[treenode] {\formatnode{6}{30}{20.0}}
                            edge from parent node[el, above] {PCR+} }
                    edge from parent node[above, pos=.4] {PA view} }
            node[above=2mm] {Total};
    \end{tikzpicture}%
    \vskip -0.1in
    \caption{Error analysis tree for the proposed model. Nodes indicate `errors / instances (ratio)', and ones highlighted in light red have error rates higher than the overall 20.0\%. `PCR+/--/?' correspond to RT-PCR positive / negative / unknown.}
    \label{fig:error_analysis}
    \centering
    \captionof{table}{Model error distribution stratified by original class and PA/AP view (`errors / instances (ratio)')}
    \label{tab:errors_class_view}
    \vskip 0.1in
    \footnotesize
    \setlength{\tabcolsep}{3pt}
    \newcolumntype{y}{r@{\,/\,}r@{ }>{\itshape\color{black!30}(}r<{\%)}}
    \begin{tabular}{@{}lyy|y@{}}
        \toprule
        Class & \multicolumn{3}{c}{PA view} & \multicolumn{3}{c}{AP view} & \multicolumn{3}{|c}{Total}\\
        \midrule
        \CVX0    & 1 & 61 & 1.6      & 7 & 39 & 17.9    & 8 & 100 & 8.0 \\
        \CVX1    & 1 & 12 & 8.3      & 6 & 89 & 6.71    & 7 & 101 & 6.9 \\
        \CVX2    & 13 & 21 & 61.9    & 20 & 77 & 26.0   & 33 & 98 & 33.7 \\
        \CVX3    & 3 & 28 & 10.7     & 29 & 73 & 39.7   & 32 & 101 & 31.7 \\
        \midrule
        Total    & 18 & 122 & 14.8   & 62 & 278 & 22.3  & 80 & 400 & 20.0 \\
        \bottomrule
    \end{tabular}
    
\end{figure}

To further investigate this effect, we analysed error patterns across classes and views (\cref{tab:errors_class_view}).
As expected, mistakes are notably more frequent in the ambiguous \CVX2 class and extremely diverse \CVX3 class.
On the other hand, \CVX0 lungs in standard PA view were the most accurately predicted by our model as well as by the annotators.
Lastly, we note that \CVX2--PA appears extremely challenging not only for the model but also for the annotators, who attained error rates of 52\%, 81\%, and 95\% for these images.
We observe similar error patterns for \textit{Ann.~1} in \cref{tab:annotator_1_errors_stratified}, and a more detailed analysis of the clinicians' performance and disagreements is presented in \cref{sec:iov_process}.


\section{Discussion}

In this study, we developed and evaluated a clinically informed hierarchical interpretation of a ML model for detecting signs of COVID-19 in chest X-rays.
By aligning with the experts' decision-making process, this formulation led to more transparent engagement with the clinical partners and helped created trust in the ML model.
Furthermore, this enabled systematic evaluation on clinically relevant predictive sub-tasks, which suggested that the model performs at least as accurately as clinicians on these challenging problems.
While not attempted here, we also envision that the operating points for each sub-task can be chosen independently, and appropriate confidence thresholds could be set for deferring decisions to human experts.

Moreover, we conducted a detailed data-driven analysis of model failures to understand in which circumstances the model's predictions may be less reliable.
Although this kind of error analysis is not often found in healthcare-related ML studies, we believe it is crucial for providing transparency and actionable insights about a model's behaviour.
For example, we may consider additional inputs to the model (here, AP/PA view) and/or complementing the training set with more data from underperforming strata.
The analysis may also be useful after deployment if presented as reliability information alongside the model's predictions.

We envisage the deployment of this model in a front-line hospital setting to automatically identify features of COVID-19 in chest X-rays. This would require validation against radiologists in a prospective multi-site study.
A successful model may potentially ease pressures on already stretched radiology services and aid less experienced clinical staff in decision-making. This can be used in conjunction with clinical status and RT-PCR in efficiently distributing patients from the front-line areas to other hospital zones thus avoiding in-hospital bottlenecks.

\section*{Acknowledgements}
This project was supported by the Microsoft Studies in Pandemic Preparedness program, with Azure credits provided by Microsoft AI for Health. The NIH-CXR and CheXpert datasets were published by the NIH Clinical Center and Stanford University School of Medicine, respectively.

We gratefully acknowledge UHB Radiology Informatics (Carol Payne), UHB Informatics (Suzy Gallier), and PIONEER for the provision of the private UHB dataset.
We would also like to thank Matthew Lungren, Kenji Takeda, and Usman Munir for the feedback and support; Omkar More, Shu Peng, and Vijay Kannan for their efforts in implementing DICOM chest X-ray support in the Azure ML labelling tool and facilitating the IOV study; as well as the study participants for their contribution.

\bibliographystyle{icml2021}
\bibliography{references}

\appendix
\numberwithin{figure}{section}
\numberwithin{table}{section}

\section{Pre-training comparison}\label{app:additional_results}

\Cref{tab:crossval_results} presents the model performance obtained in the classification tasks illustrated in \cref{fig:model_arch} in terms of precision-recall area-under-the-curve (PR-AUC), ROC-AUC, and accuracy metrics, along with multi-class (one-of-four) accuracy, in 5-fold cross-validation runs. 

The experimental results demonstrate that model pre-training significantly improves the classification accuracy for every task in the hierarchical decision making process. The same behaviour is observed when we pre-trained the models on two different external datasets: NIH-CXR \cite{wang2017nih} and CheXpert \cite{irvin2019chexpert}. However, there was no significant difference between the compared pre-trained models whilst they consistently outperformed models trained with random weight initialisation. 
Due to comparable performance of the pre-trained models, we used NIH pre-training for the remaining experiments.

\begin{table}[htb]
    \centering
    \caption{Cross-validation classification results aggregated over $K=5$ folds ($N=4940$; \CVX0:~1154, \CVX1:~1778, \CVX2:~1093, \CVX3:~915). Mean ± (std.).}
    \label{tab:crossval_results}
    \vskip 0.1in
    \setlength{\tabcolsep}{3pt}
    \resizebox{\linewidth}{!}{\begin{tabular}{@{}llccc@{}}
        \toprule
        Pre-training & Classification task & PR-AUC      & ROC-AUC	    & Accuracy \\
        \midrule
        None        & \Sugg.\ COVID & .922 ± .005   & .893 ± .006   & .815 ± .012 \\
                    & \CVX1 vs \CVX2& .722 ± .023   & .835 ± .014   & .758 ± .013 \\
                    & \CVX0 vs \CVX3& .793 ± .043   & .843 ± .029   & .772 ± .033 \\
                    & Multi-class   & --- 	        & ---           & .625 ± .018 \\
        \midrule
        NIH-CXR     & \Sugg.\ COVID & .938 ± .007   & .915 ± .009   & .837 ± .009 \\
                    & \CVX1 vs \CVX2& .746 ± .023   & .857 ± .005   & .777 ± .012 \\
                    & \CVX0 vs \CVX3& .813 ± .031   & .855 ± .017   & .784 ± .016 \\
                    & Multi-class   & --- 	        & ---           & .655 ± .016 \\
        \midrule
        CheXpert    & Multi-class   & --- 	        & ---           & .653 ± .015 \\
        \midrule
        Both        & Multi-class   & --- 	        & ---           & .658 ± .015 \\
        \bottomrule
    \end{tabular}}
\end{table}


\section{Implementation details}
\label{sec:implementation_details}

Our model uses a DenseNet-121 \citep{huang2017densenet} backbone for image feature extraction. The model is trained using cross entropy loss over the 4 classes for 50 epochs with a batch size of 64. We use the Adam optimiser \citep{kingma2015adam} and a learning rate of $10^{-5}$. The pixel values of each image are linearly normalised between $0$ and $255$. During training and inference, each image is resized to size $256{\times}256$, and then a center crop taken to get an image of size $224{\times}224$. When training, we perform data augmentation using random horizontal flips, affine transforms, random crops, and brightness, contrast, saturation and gamma transforms.

\section{Multi-expert labelling process}\label{sec:iov_process}
In this study, we formed an isolated test set with multiple expert labels to assess the clinical applicability of the learnt models. For this purpose, a panel of three clinicians manually labelled a subset of the in-house dataset ($N=400$) described in \cref{sec:multi_label_dataset}. The labelling process was carried out using a cloud-based annotation tool on an image-by-image basis, where each expert was able to adjust the image intensity window/level and analyse CXR scans in high-resolution. At the same time, the annotation time of each expert was monitored throughout this exercise to create a proxy measure quantifying the difficulty of each labelling task and also the experience level of each annotator.

\begin{figure}[htb]
    \centering
    \includegraphics[width=.95\linewidth]{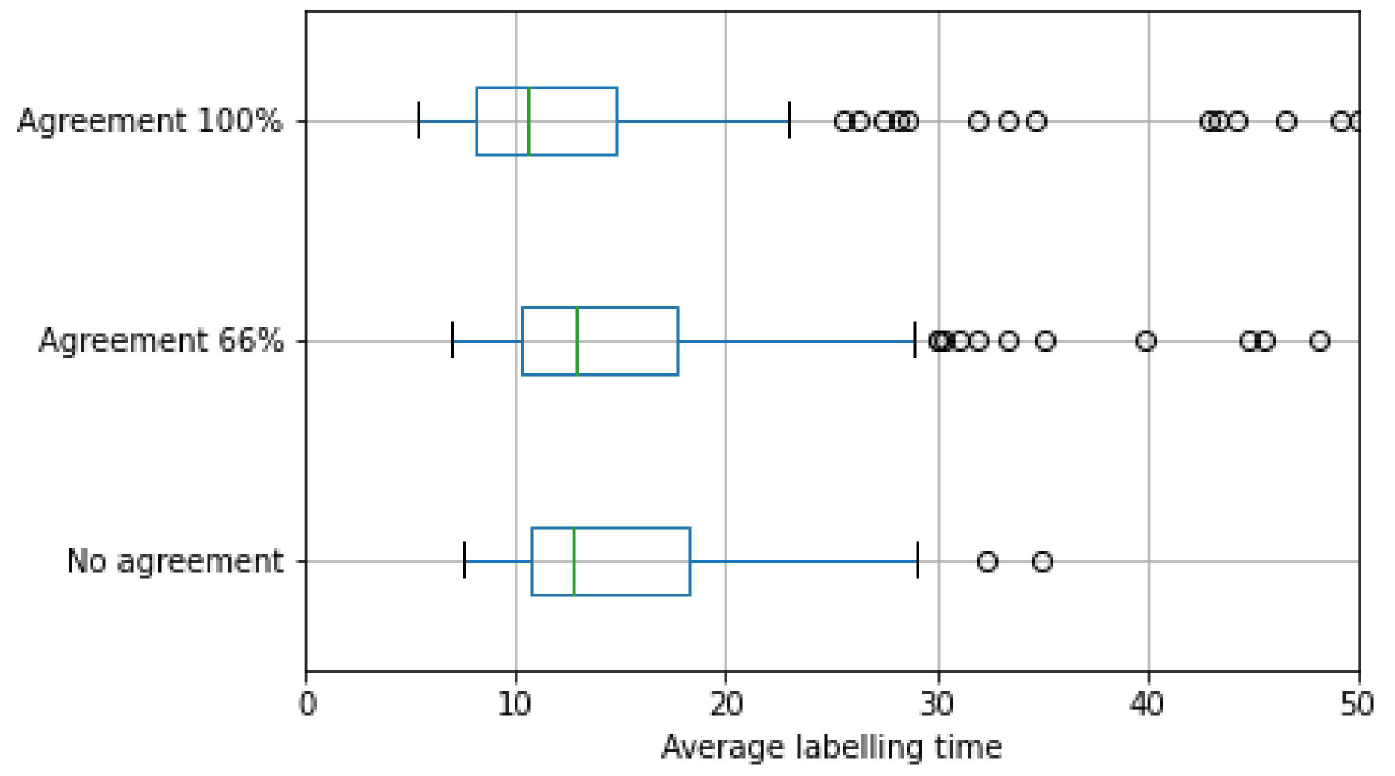}
    \vskip -0.1in
    \caption{Distribution of average labelling time (in seconds) on 380 CXR images, broken down by how often the three clinicians agreed on the class assignment. A small subset of of the 400 images ($N=11$) whose average labelling time exceeded 50 seconds were omitted from the graph to reduce the noise level in the analysis, and a further 9 were skipped by at least one annotator.}
    \label{fig:labelling_time_boxplot}
\end{figure}

\Cref{fig:labelling_time_boxplot} shows a breakdown of average labelling time by agreement of the 3 annotators. We observe that, for the samples where the experts arrive at the same conclusion, the annotation time is consistently lower compared to the (presumably more difficult) samples where they do not agree.
Additionally, we observed a systematic correlation between the annotators' disagreement and the model's errors (see \cref{tab:disagreement_vs_erros}), suggesting that the model tended to have more difficulty on ambiguous cases.
Lastly, \cref{tab:annotator_1_errors_stratified} reports the breakdown of errors of the most experienced annotator (\textit{Ann.~1}) by class label and CXR view. Comparing to \cref{tab:errors_class_view}, it suggests that the model and \textit{Ann.~1} have similar error patterns for this task.

\begin{table}[htb]
    \centering
    \caption{Model error rates versus expert disagreement for each predictive task. We indicate agreement patterns as `3' for full agreement, `2:1' for partial agreement, and `1:1:1' for full disagreement between the three annotators. The model's overall error rates are also included for reference.}
    \label{tab:disagreement_vs_erros}
    \vskip 0.1in
    \footnotesize
    \begin{tabular}{@{}lcccc@{}}
        \toprule
        & \multicolumn{3}{c}{Annotator agreement} \\
        \cmidrule{2-4}
        Classification task & 3 & 2:1       & 1:1:1     & Overall \\
        \midrule
        \Sugg.\ COVID & 18.6\%  & 22.6\%    & ---       & 20.0\% \\
        \CVX1 vs \CVX2& 19.5\%  & 33.9\%    & ---       & 27.6\% \\
        \CVX0 vs \CVX3& 16.4\%  & 29.9\%    & ---       & 20.9\% \\
        Multi-class   & 30.3\%  & 47.9\%    & 56.1\%    & 40.8\% \\
        \bottomrule
    \end{tabular}
\end{table}

\begin{table}[htb]
    \centering
    \captionof{table}{Distribution of disagreements between the labels provided by a consultant chest radiologist (\textit{Ann.~1}) and the reference labels collected on the test set. The results are stratified by target class and PA/AP view (`errors / instances (ratio)').}
    \label{tab:annotator_1_errors_stratified}
    \vskip 0.1in
    \footnotesize
    \setlength{\tabcolsep}{3pt}
    \newcolumntype{y}{r@{\,/\,}r@{ }>{\itshape\color{black!30}(}r<{\%)}}
    \begin{tabular}{@{}lyy|y@{}}
        \toprule
        Class & \multicolumn{3}{c}{PA view} & \multicolumn{3}{c}{AP view} & \multicolumn{3}{|c}{Total}\\
        \midrule
        \CVX0    & 1 & 61 & 1.6      & 6 & 39 & 15.4    & 7 & 100 & 7.0 \\
        \CVX1    & 1 & 12 & 8.3      & 17 & 89 & 19.1    & 18 & 101 & 17.8 \\
        \CVX2    & 11 & 21 & 52.4    & 34 & 77 & 44.2   & 45 & 98 & 45.9 \\
        \CVX3    & 3 & 28 & 10.7     & 17 & 73 & 23.3   & 20 & 101 & 19.8 \\
        \midrule
        Total    & 16 & 122 & 13.1   & 74 & 278 & 26.6  & 90 & 400 & 22.5 \\
        \bottomrule
    \end{tabular}
\end{table}

\end{document}